\documentclass[preprint2]{aastex}

\usepackage[]{natbib}
\usepackage{graphicx}
\usepackage{amssymb}
\usepackage{color}
\usepackage{ulem}

\newcommand{\etacar}{$\eta$~Car}
\newcommand{\ess}{EHG7}

\newcommand{\ASCA}{{\it ASCA}}

\newcommand{\EINSTEIN}{{\it Einstein}}

\newcommand{\ROSAT}{{\it ROSAT}}
\newcommand{\XMM}{{\it XMM-Newton}}

\newcommand{\CHANDRA}{{\it Chandra}}

\newcommand{\UNITFLUX}{{\rm ergs~cm$^{-2}$~s$^{-1}$}}

\newcommand{\UNITLUMI}{{\rm ergs~s$^{-1}$}}

\newcommand{\UNITNH}{{\rm cm$^{-2}$}}
\newcommand{\UNITVEL}{{\rm km~s$^{-1}$}}

\newcommand{\UNITSOLARMASS}{{\it M$_{\odot}$}}

\newcommand{\DEGREE}{{$^{\circ}$}}
\newcommand{\ARCMIN}{{$'$}}

\newcommand{\ARCSEC}{{$''$}}

\newcommand{\NH}{{\it N$_{\rm H}$}}

\newcommand{\LX}{{\it L$_{\rm X}$}}

\newcommand{\KT}{{\it kT}}

\slugcomment{Accepted for publication to ApJL}

\shorttitle{Carina East Source}
\shortauthors{Hamaguchi et al.}

\begin{document}

\title{A Smoking Gun in the Carina Nebula}

\author{Kenji Hamaguchi\altaffilmark{1,2}, Michael F. Corcoran\altaffilmark{1,3}, Yuichiro Ezoe\altaffilmark{4}, 
Leisa Townsley\altaffilmark{5}, Patrick Broos\altaffilmark{5}, Robert Gruendl\altaffilmark{6}, Kaushar Vaidya\altaffilmark{6,a},
Stephen M. White\altaffilmark{7}, Tod Strohmayer\altaffilmark{8}, Rob Petre\altaffilmark{8}, You-Hua Chu\altaffilmark{6}}

\altaffiltext{1}{CRESST and X-ray Astrophysics Laboratory NASA/GSFC, Greenbelt, MD 20771}
\altaffiltext{2}{Department of Physics, University of Maryland, Baltimore County, 1000 Hilltop Circle, Baltimore, MD 21250}
\altaffiltext{3}{Universities Space Research Association, 10211 Wincopin Circle, Suite 500, Columbia, MD 21044}
\altaffiltext{4}{Tokyo Metropolitan University, 1-1, Minami-Osawa, Hachioji, Tokyo, 192-0397, Japan}
\altaffiltext{5}{Department of Astronomy and Astrophysics, Pennsylvania State University, 525 Davey Laboratory, University Park, PA 16802}
\altaffiltext{6}{Department of Astronomy, University of Illinois, Urbana, IL 61801}
\altaffiltext{7}{Department of Astronomy, University of Maryland, College Park, MD 20742}
\altaffiltext{8}{Astrophysics Science Division, NASA Goddard Space Flight Center, Greenbelt, MD 20771}
\altaffiltext{a}{previously Kaushar Sanchawala}

\begin{abstract}
The Carina Nebula is one of the youngest, most active sites of massive star formation in our Galaxy.
In this nebula, we have discovered a bright X-ray source that has persisted for $\sim$30 years.
The soft X-ray spectrum, consistent with \KT~$\sim$128~eV blackbody radiation with mild extinction,
and no counterpart in the near- and mid-infrared wavelengths indicate that it is a $\sim$10$^{6}$-year-old neutron 
star housed in the Carina Nebula.
Current star formation theory does not suggest that
the progenitor of the neutron star and massive stars in the Carina Nebula, in particular \etacar, are coeval.
This result suggests that the Carina Nebula experienced at least two major episodes of massive star formation.
The neutron star may be responsible for remnants of high energy activity seen in multiple wavelengths.
\end{abstract}


\section{Introduction}

Massive stars ($M \gtrsim10$\UNITSOLARMASS) are born from giant molecular clouds
along with many lower mass stars, forming a stellar cluster or association.
Massive stars evolve orders of magnitude  more quickly than lower mass stars,
and die through supernova or hypernova explosions in $\lesssim$10$^{7}$ years.
High interstellar pressures that massive stars produce through their strong UV radiation and 
supernova explosions can compress a pre-existing cloud and trigger the formation of new stars \citep[cf.][]{Elmegreen1998}.

The Carina Nebula is one of the most massive star forming regions in our Galaxy.
It contains two massive stellar clusters, Trumpler 14 and 16 (Figure 1),
possessing over 50 massive stars with spectral types earlier than O6
\citep[$\gtrsim$40~\UNITSOLARMASS,][]{Smith2006}.
The nebula is also home to one of the most massive stars in our Galaxy, $\eta$ Carinae,
which has an estimated initial mass $\gtrsim$150~\UNITSOLARMASS\ and a current mass of about 90~\UNITSOLARMASS\ \citep{Hillier2001}.

At a distance of only $\sim$2.3~kpc \citep{Davidson1997},
the Carina Nebula is one of the best sites for studying how very massive stars form and affect their environment.
It shows signatures of violent activities:
a bipolar supershell structure \citep{Smith2000}, strong turbulence in interstellar clouds
\citep{Yonekura2005}, and hot X-ray plasma through the entire nebula
\citep{Seward1979,Townsley2006,Hamaguchi2007a,Ezoe2008}.
Two primary mechanisms have been proposed to produce these structures.
One is strong winds and UV radiation from massive stars in the nebula, while the other is
supernova explosions.
The energy budget and elemental abundance distribution favor the supernova mechanism \citep{Yonekura2005,Hamaguchi2007a}.
However, neither black hole, neutron star, nor clear remnant from a supernova,
has been found in the Carina Nebula.
In addition, any massive supernova progenitor is unlikely to be co-eval with the observed stars since it would be more massive than \etacar; 
this means that, if a supernova occurred in the Carina Nebula, its progenitor would probably have formed much earlier than \etacar. Thus, detection of any compact object in the Carina Nebula would be of vital interest
for understanding the star formation history of the region.

From multiple sets of X-ray data,
we found a promising neutron star candidate at the heart of the Carina Nebula.
The source, which we call \ess\ \citep{Ezoe2008}, was detected as a very soft source in X-ray images 8.5\ARCMIN\ southeast of \etacar,
equivalent to 5.7~pc in the projected physical distance assuming $d =$2.3~kpc (See Figure~1).
The X-ray source has also been reported contemporaneously in four other papers \citep{Colombo2003,Ezoe2008,Pires2008,Pires2008b},
the last two also suggested \ess\ as a neutron star based on \XMM\ and \CHANDRA\ data.
This paper presents a comprehensive study of the characteristics of \ess\ using all available X-ray data 
from \EINSTEIN\ and \ROSAT\ observations to the latest \XMM\ observations in 2009
to present conclusive evidence of the compact nature of \ess.

\section{X-ray Observations and Analysis}

We found 32 X-ray observations with \ess\ in the field of view and spatial resolution better than 1\ARCMIN\
in the HEASARC archive and our proprietary data (Table~\ref{tbl:obslogs}).
Observations with lower spatial resolution (e.g. \EINSTEIN\ IPC, \ASCA)
failed to detect \ess\ due to severe contamination from surrounding soft diffuse X-ray emission.
Throughout this document, individual observations with \EINSTEIN, \ROSAT, \XMM, and \CHANDRA\
observations are designated EIN, ROS, XMM, CXO respectively, subscripted 
with the year, month, and day of the start time of the observations.
 
We used the software package SAS, version 8.0.1 for analysis of the \XMM\ EPIC data
\citep{Struder2001}.
We followed the standard method\footnote{http://xmm.esac.esa.int/sas/8.0.0/documentation/threads/}
in data processing --- screening out high background periods, selecting photon events and generating
spectral response files.
For the timing and spectral analysis, we took source events from an encircled region with 30\ARCSEC\ radius and
background events from an annulus region with 60\ARCSEC\ outer and 30\ARCSEC\ inner radii.
For data sets with the source close to 2 CCD gaps (see Table~\ref{tbl:obslogs}),
source events are collected only from a CCD chip with the source peak.
Comparison of these spectra with spectra of the other cameras in the same observation looked consistent,
but we did not use them for flux measurements.

\begin{deluxetable}{lclcccc}
\tablecolumns{7}
\tablewidth{0pc}
\tabletypesize{\scriptsize}
\tablecaption{X-ray Observations with \ess\ in the Field of View \label{tbl:obslogs}}
\tablehead{
\colhead{Abbreviation}&
\colhead{Observation ID}&
\colhead{Observation Start}&
\colhead{Detector/Filter}&
\colhead{Offset}&
\colhead{Exposure}&
\colhead{Flux (0.3$-$2~keV)}
\\
&&&&\colhead{(arcmin)}&\colhead{(ksec)}&\colhead{(10$^{-13}$~\UNITFLUX)}
}
\startdata
EIN$_{\rm 781214}$&1074&1978 Dec. 14, 05:44&HRI-2&8.5&17.4&0.89$\pm$0.54\\
ROS$_{\rm 900727}$&CA150037H.N1&1990 Jul. 27, 00:12&HRI&15.0&3.3&$<$1.1\\
ROS$_{\rm 911215}$&US200108P.N1&1991 Dec. 15, 9:58&PSPCB&19.4&1.6&$<$2.2\\
ROS$_{\rm 920612}$&US900176P.N1&1992 Jun. 12, 22:33&PSPCB&8.7&23.6&0.94$\pm$0.16\\
ROS$_{\rm 920731}$&WG900385H.N1&1992 Jul. 31, 01:03&HRI&8.7&11.4&1.04$\pm$0.33\\
ROS$_{\rm 920809}$&WG201262P.N1&1992 Aug. 9, 22:21&PSPCB&13.8&5.7&1.43$\pm$0.47\\
ROS$_{\rm 920810}$&WG200709P-1.N1&1992 Aug. 10, 19:03&PSPCB&35.4&5.9&$<$2.4\\
ROS$_{\rm 921215}$&US900176P-1.N1&1992 Dec. 15, 17:39&PSPCB&8.7&14.1&0.69$\pm$0.18\\
ROS$_{\rm 940106}$&WG900385H-2.N1&1994 Jan 6, 1:54&HRI&8.7&0.5&$<$7.7\\
ROS$_{\rm 940721}$&WG900385H-3.N1&1994 Jul. 21, 02:04&HRI&8.7&40.1&0.91$\pm$0.18\\
ROS$_{\rm 960813}$&US900644H.N1&1996 Aug. 13, 21:07&HRI&8.7&1.7&$<$3.3\\
ROS$_{\rm 971223}$&US202331H.N1&1997 Dec. 23, 08:59&HRI&8.7&46.5&0.82$\pm$0.16\\
XMM$_{\rm 000726}$&112580601&2000 Jul. 26, 05:08&MOS/thick&8.7&--/33.2/30.1$^{\dagger}$&1.04$\pm$0.18\\
XMM$_{\rm 000727}$&112580701&2000 Jul. 27, 23:58&MOS/thick&8.7&--/10.9/7.9$^{\dagger}$&1.21$\pm$0.30\\
XMM$_{\rm 030125}$&145740101&2003 Jan. 25, 12:58&MOS/thick&8.4&--/6.9/6.9$^{\dagger}$&1.40$\pm$0.36\\
XMM$_{\rm 030127A}$&145740201&2003 Jan. 27, 01:04&MOS/thick&8.4&--/6.8$^{\dagger}$/6.9$^{\dagger}$&(1.21$\pm$0.25)\\
XMM$_{\rm 030127B}$&145740301&2003 Jan. 27, 20:37&MOS/thick&8.4&--/6.8$^{\dagger}$/6.8$^{\dagger}$&(1.11$\pm$0.23)\\
XMM$_{\rm 030129A}$&145740401&2003 Jan. 29, 01:41&MOS/thick&8.4&--/8.4$^{\dagger}$/8.4&0.95$\pm$0.27\\
XMM$_{\rm 030129B}$&145740501&2003 Jan. 29, 23:55&MOS/thick&8.4&--/6.9$^{\dagger}$/6.9&1.30$\pm$0.32\\
XMM$_{\rm 030608}$&160160101&2003 Jun. 08, 13:30&MOS/thick&8.4&--/17.0/16.5&1.10$\pm$0.19\\
XMM$_{\rm 030613}$&160160901&2003 Jun. 13, 23:52&MOS/thick&8.4&--/31.1/31.1&1.03$\pm$0.15\\
XMM$_{\rm 030722}$&145780101&2003 Jul. 22, 01:51&MOS/thick&8.4&--/8.4/8.4$^{\dagger}$&1.19$\pm$0.34\\
XMM$_{\rm 030802}$&160560101&2003 Aug. 02, 21:01&MOS2/thick&8.4&--/--/11.9&1.14$\pm$0.26\\
XMM$_{\rm 030809}$&160560201&2003 Aug. 09, 01:44&MOS/thick&8.4&--/12.2/12.2&1.06$\pm$0.20\\
XMM$_{\rm 030818}$&160560301&2003 Aug. 18, 15:23&MOS/thick&8.4&--/18.5/18.5&1.06$\pm$0.17\\
XMM$_{\rm 041207}$&206010101&2004 Dec. 7, 7:30&pn/medium&16.3&19.4$^{\dagger}$/--/--&(0.88$\pm$0.14)\\
XMM$_{\rm 060131}$&311990101&2006 Jan. 31, 18:04&pn\&MOS/thick&8.7&25.2/65.3$^{\dagger}$/65.3&1.05$\pm$0.12\\
CXO$_{\rm 080905}$&9488&2008 Sep. 5, 21:24&ACIS-I&6.7&59.4&0.87$\pm$0.08\\
XMM$_{\rm 090105}$&560580101&2009 Jan 5, 10:23&MOS/thick&8.5&--/14.0/13.9&1.17$\pm$0.19\\
XMM$_{\rm 090109}$&560580201&2009 Jan 9, 14:28&MOS/thick&8.5&--/11.4/11.4&1.17$\pm$0.21\\
XMM$_{\rm 090115}$&560580301&2009 Jan 15, 11:23&MOS/thick&8.5&--/25.4/25.4&0.97$\pm$0.15\\
XMM$_{\rm 090202}$&560580401&2009 Feb 2, 04:46&MOS2/thick&8.5&--/--/26.2&1.09$\pm$0.18\\
\enddata
\tablecomments{
Abbreviation: EIN: Einstein, ROS: ROSAT, XMM: XMM-Newton, CXO: Chandra, 
Detector/Filter: optical blocking filter for \XMM,
Offset: angle of \ess\ from the nominal coordinate,
Exposure: pn/MOS1/MOS2 for \XMM,
Flux: Error denotes the 90\% confidence range. The upper limit is at a 3$\sigma$ level.
The flux errors include uncertainty of the absolute flux calibration of each telescope 
(\EINSTEIN, \ROSAT, \XMM: 10\%, \CHANDRA: 5\%, 
ROSAT User's Handbook --- http://heasarc.gsfc.nasa.gov/docs/rosat/ruh/handbook/handbook.html, 
\citet{Beuermann2006}, 
XMM-SOC-CAL-TN-0018 --- http://xmm2.esac.esa.int/external/xmm\_sw\_cal/calib/documentation/index.shtml\#EPIC, 
Calibration Requirements and Present Status --- http://asc.harvard.edu/cal/docs/cal\_present\_status.html\#abs\_eff).
$^{\dagger}$the source peak was close to 2 CCD gaps or an edge of the field of view.
Those data sets were not used for flux measurements in the right column.
}
\end{deluxetable}

\begin{figure*}[t]
\epsscale{2.1}
\plotone{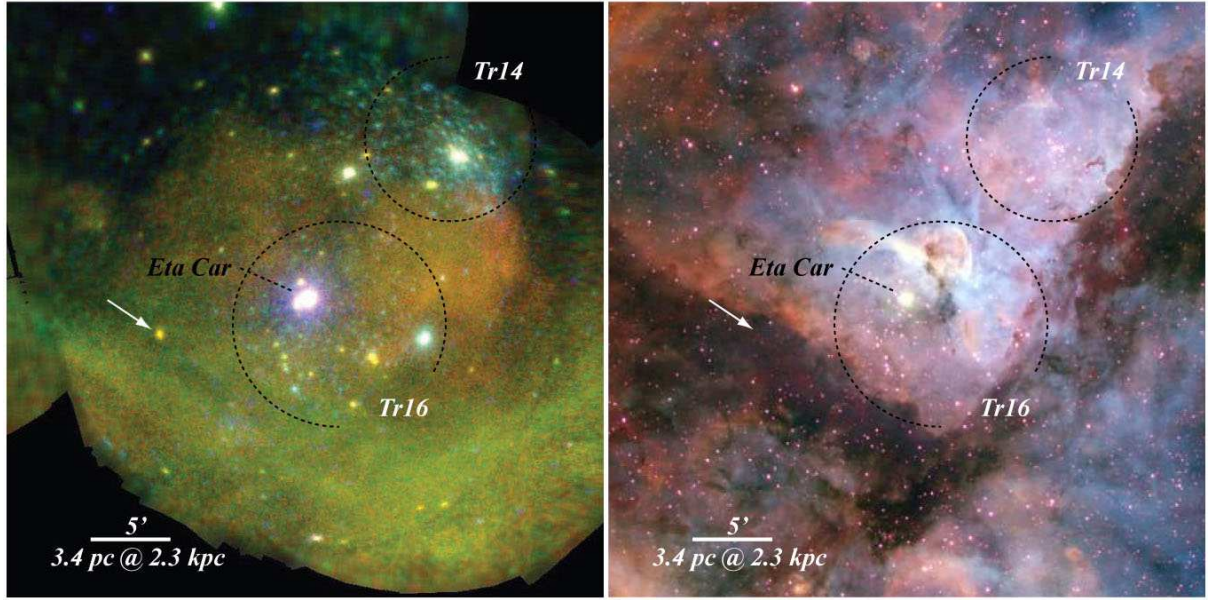}
\caption{
The central part of the Carina Nebula in X-rays 
taken with \XMM\ ({\it Left}, red: 0.4$-$0.7~keV, green: 0.7$-$1.3~keV, blue: 2$-$7~keV)
and in optical emission lines ({\it Right}, Credit: Nathan Smith, University of Minnesota/NOAO/AURA/NSF).
\ess, the source discussed here, is denoted by the white arrow.  
It is located on the conspicuous V-shaped dark lane running across the optical image 
and is surrounded by diffuse X-ray emission.
The black dotted line and circles depict positions of the super massive star \etacar\ 
and massive stellar clusters, Trumpler 14 and 16, respectively.
\label{fig:imgxrayoptical}}
\end{figure*}

For source position determination on the \CHANDRA\ image,
we used a custom analysis method for \CHANDRA\ data sets developed by \citet{Broos2007} and Broos et al. (in prep).
In producing the \CHANDRA\ spectral and timing data sets, 
we analyzed level 2 event data using the software package CIAO, version 4.0.
We took source events from an encircled region with a 8\ARCSEC\ radius and
background events from an annulus region with 16\ARCSEC\ outer and 8\ARCSEC\ inner radii.

For general light curve and spectral analysis of the above data sets and for the \ROSAT\ and \EINSTEIN\ data processing,
we used the software package HEAsoft version 6.5.1.
We started from the pipeline products of the \ROSAT\ and \EINSTEIN\ data.
For the \ROSAT\ image analysis,
we excluded PHA channels above 10 to reduce particle background events for the HRI
and extracted events between 0.3$-$2~keV for the PSPC.
For the PSPC spectral and timing analysis,
we took source events from an encircled region with 30\ARCSEC\ radius and
background events from an annulus region with 60\ARCSEC\ outer and 30\ARCSEC\ inner radii.
For \EINSTEIN\ HRI calibration information, 
we referred the mirror vignetting at the location of \ess\ in EIN$_{\rm 781224}$ (8.5\ARCMIN\ off-axis) 
to \citet{Harnden1984} (91\% at 0.6~keV)
and estimated encircled energy of the point spread function (PSF) within a 32\ARCSEC$\times$32\ARCSEC\ box at 76\%
using a calibration observation of Cyg X-1 at a similar off-axis angle (10\ARCMIN: the dataset H1956N35.xic).
The dead time was negligible and therefore ignored.

\begin{figure}[h]
\epsscale{.95}
\plotone{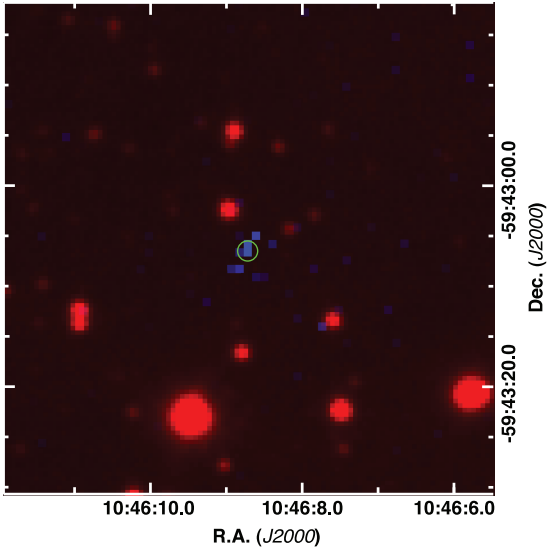}
\caption{Magnified image of the neutron star field ({\it blue}: a \CHANDRA\ X-ray image after the {\it PSF} image reconstruction,
{\it red}: a deep $H$ band image taken with the IRSF observatory).
The position of the X-ray source \ess\ is marked with a green circle with a 1\ARCSEC\ radius; it has no infrared counterpart.
Another X-ray source about 10\ARCSEC\ southwest from \ess\ has a counterpart in the $H$-band.
\label{fig:imgmag}}
\end{figure}

\section{Results}

The \CHANDRA\ observation CXO$_{\rm 080905}$, performed as a part of a Very Large Project to
map the Carina Nebula in X-rays (PI: Leisa Townsley), gives the most precise and reliable absolute position of \ess~(Figure~\ref{fig:imgmag}).
In this observation, \ess\ was imaged 6.7\ARCMIN\ off-axis on the ACIS-I detector.
The absolute coordinates of the source were determined to be
($\alpha_{2000}$, $\delta_{2000}$) = (10$^{\rm h}$46$^{\rm m}$08.72$^{\rm s}$, $-$59\DEGREE43\ARCMIN06.5\ARCSEC)
with less than 1\ARCSEC\ positional uncertainty after correcting the Chandra astrometry by cross correlation of 
X-ray and infrared sources from the 2MASS catalogue and correlating the PSF at the location of this source 
with the image of the source.

We searched for near- and mid-infrared counterparts from images obtained with the IRSF (InfraRed Survey Facility) 
1.4m telescopes \citep[see][for details]{Sanchawala2007a,Sanchawala2007b}
and in all available {\it Spitzer} InfraRed Array Camera \citep[IRAC;][]{Fazio2004} observations combined,
but these images showed no counterpart object within the error circle of the X-ray source
(Figure~\ref{fig:imgmag}).
We used standard aperture photometry routines in IRAF to estimate the 3-$\sigma$ upper-limits,
which were 18.5~mag, 19.5~mag and 18.5~mag in the $J, H$ and $K_{S}$ bands and 
0.84, 0.51, 4.7, and 13.6~mJy in the IRAC 3.6, 4.5, 5.8, and 8.0~$\mu$m bands, respectively.
Assuming blackbody spectra with temperatures between 2000 and 10000~$K$, 
the bolometric flux of \ess\ was limited to $<$5$\times$10$^{-13}$~\UNITFLUX.
We note that \citet{Pires2008b} measured upper limits
of 25$^{m}$ in $R$ and $B$-band optical images.

X-ray emission from \ess\ was detected at above 3$\sigma$ significance in all observations in Table~\ref{tbl:obslogs}
except for 5 \ROSAT\ observations with short exposures.
None of these observations showed any significant time variation from \ess\ at greater than 90\% confidence
in a fit of each light curve in the whole instrumental energy band with a constant flux model.
For data sets with no spectral resolution and/or poor photon statistics,
we measured the photon count rate or upper limit in whole band images using the {\tt sosta} package in {\tt ximage},
taking a suitable nearby source free region as background.
We then converted the source counts to energy flux using the {\tt PIMMS} tool,
assuming blackbody radiation with \KT~$\sim$128~eV and \NH~$\sim$3.2$\times$10$^{21}$~\UNITNH,
derived from a best-fit model to the \XMM\ spectrum as described below.
For the other data sets, we measured source flux from time averaged spectra, assuming the same spectral model.
The results are shown in Table~\ref{tbl:obslogs}.
\ess\ did not show any significant flux variation above $\sim$30\% between the observations
though a formal fit of these data rejected a constant flux model at above 90\% confidence ($\chi^{2}$/d.o.f = 38.45/23).
The X-ray emission was stable for $\sim$30 years with $F_{\rm X}$ (0.3$-$2~keV) $\approx$1.06$\times$10$^{-13}$ \UNITFLUX.

\begin{figure}[h]
\epsscale{1.0}
\plotone{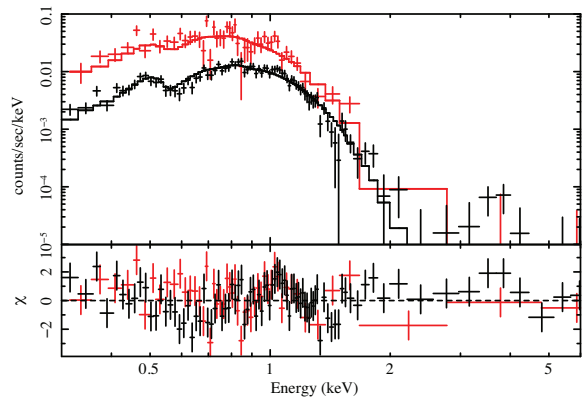}
\caption{
{\it Top panel} : \XMM\ MOS ({\it black}) and pn ({\it red}) spectra combined from all the available data.
The solid line shows the best-fit model of the spectra between 0.3$-$3 keV by an absorbed blackbody model.
{\it Bottom panel} : the residuals of the $\chi^{2}$ fit.
\label{fig:specxmmmos}}
\end{figure}

Individual observations did not have sufficient photon statistics for detailed spectral analysis.
Since the spectral shape did not change significantly between observations\footnote{We simultaneously 
fitted 0.3$-$3~keV spectra of all the available EPIC cameras in each observation by an absorbed blackbody model 
and found that 95\% confidence ranges of the \NH\ and \KT\ parameters overlapped with each other.},
we combined all the source spectra taken between 2000 and 2006 
with the \XMM\ MOS and pn detectors (Figure~\ref{fig:specxmmmos}).
We used the HEAsoft tool {\tt mathpha} to combine the source and background spectra, 
{\tt marfrmf} to merge {\tt arf} and {\tt rmf} files and then {\tt addrmf} to average produced response files over
by weighting them by net source counts.
The combined X-ray spectrum showed significant emission only below 2~keV and is
almost featureless except for a slight dip at around 0.9~keV and a strong dip at around 0.6~keV.
The latter feature at $\sim$0.6~keV was perhaps produced by edge absorption of oxygen in the interstellar medium
and in the CCD detector response.
The spectrum can be reproduced using a 1-temperature optically-thin thermal plasma emission model
suffering absorption by neutral gas along the line of sight (Table~\ref{tbl:specfit}),
but the best-fit model has an unusually small elemental abundance of $\lesssim$10$^{-3}$ solar,
without evidence of emission lines from oxygen, iron, or neon atoms near 0.5$-$1~keV.
The spectrum also can be fit by a 1-temperature blackbody emission model with a temperature of
128$\pm$7~eV (90\% confidence) suffering absorption by neutral gas with hydrogen column density of
3.2$\pm0.4\times$10$^{21}$~\UNITNH\ (90\% confidence).
The spectrum could not be adequately fit using an absorbed power law model.
We note that \citet{Pires2008b} derived a similar best-fit result in individual fits of \XMM\ spectra.

\begin{deluxetable}{lclccc}
\tablecolumns{6}
\tablewidth{0pc}
\tabletypesize{\scriptsize}
\tablecaption{Spectral Fits\label{tbl:specfit}}
\tablehead{
\colhead{Model}&
\colhead{\KT}&
\colhead{Abundance}&
\colhead{\NH}&
\colhead{\LX}&
\colhead{$\Delta\chi^{2}$ (d.o.f)}\\
&\colhead{(eV)}&\colhead{(solar)}&\colhead{(10$^{21}$\UNITNH)}&\colhead{(10$^{32}$\UNITLUMI)}
}
\startdata
1T APEC&171~(162$-$174)&$<$8.0$\times$10$^{-4}$&5.4~(5.0$-$5.8)&32&1.51~(125)\\
1T Blackbody&128~(121$-$133)&\nodata&3.2~(2.8$-$3.6)&4.6&1.49~(126)\\
\enddata
\tablecomments{
Absorption corrected \LX\ between 0.3$-$8~keV assuming $d$ =2.3~kpc.
The parentheses in the \KT\ and \NH\ columns denote the 90\% confidence range.
}
\end{deluxetable}

The hydrogen column density is consistent with interstellar absorption to the Carina Nebula
\citep[$\sim$3$\times$10$^{21}$~\UNITNH, see discussion in][]{Leutenegger2003}.
The 90\% confidence upper limit (3.6$\times$10$^{21}$~\UNITNH) is much smaller
than absorption through our Galaxy \citep[$\sim$1.2$\times$10$^{22}$~\UNITNH
,][]{Kalberla2005},
which emission from an AGN should suffer.
It is also significantly smaller than absorption to the background stars in the field, 
which have visual extinctions twice as large as Carina cluster members \citep{Degioia2001}.
This result suggests that \ess\ is in front of the molecular cloud that
lies behind the Trumpler 14 and 16 clusters,
and neighbors the Carina Nebula.

The ratio of the X-ray flux to the bolometric flux, log $F_{\rm X}$/$F_{\rm bol} >-$0.7, was significantly 
larger than that typically seen in stellar X-ray emission \cite[$\lesssim-$3,][]{Gagne1995}.
This large $F_{\rm X}$/$F_{\rm bol}$
means that \ess\ is unlikely to be a star in the Carina star-forming complex nor in the fore- and background
but must be a compact object such as a black hole, a white dwarf or a neutron star.
The X-ray characteristics of this source, a single temperature blackbody spectrum without time variation longer than hourly timescales,
are typical of isolated neutron stars, where the emission comes from the cooling neutron star surface.
It is unlikely to be a black hole,
which would show strong time variation on both short and long timescales,
reflecting activity around the accretion disk near the event horizon.
It is also unlikely to be a white dwarf, which has
normally significantly lower blackbody temperatures than 100~eV \citep{Kahabka1997}.
Although some white dwarfs with masses close to the Chandrasekhar limit can have \KT~$\gtrsim$100~eV,
they should be more than five orders of magnitudes more luminous \citep{Sala2005}.
All this evidence shows that the source is most likely a neutron star.

We found no radio counterpart of \ess\ in a continuum radio map of the Carina region from the Southern 
Galactic Plane Survey \citep{McClure2005}, but the resolution of these data is only
1\ARCMIN\ and weak point sources will not be detected against the strong nebular radio sources Car I and Car II.
The ATNF Pulsar Catalogue \citep{Manchester2005}
lists 11 pulsars within 2\DEGREE\ of \ess, but no source coincident with \ess\ itself.
We searched for a pulse from the \XMM\ pn data between 0.3$-$2~keV in XMM$_{\rm 060131}$
using the Fast Fourier Transform and  Z$^{2}$ searches.
However, we detected no significant pulses from \ess\ in the X-ray data in the frequency range between 10$^{-4}$--13.6~Hz.
A 90\% confidence upper limit on any signal amplitude in this frequency range was 27.7\% (rms).

\section{Discussion}

The intrinsic X-ray luminosity is 4.6$\times$10$^{32}$~\UNITLUMI\ or equivalent to an emitting radius of $\sim$7.3~km
assuming \KT~=128 eV and $d =$2.3~kpc.
Assuming the standard cooling curve of a $M =$1.4 \UNITSOLARMASS\ neutron star \citep{Tsuruta2006},
the age of \ess\ is estimated at $\sim$0.5$-$1$\times$10$^{6}$-year.
The blackbody temperature, 0.13~keV, is also typical of neutron stars older than 1000 years
\citep[e.g. 1E1207.4-5209, PSR B0656+14 and PSR B1055-52, ][]{Sanwal2002, Luca2005}.
The X-ray spectrum did not show a non-thermal power-law component in the hard band that would typically originate 
from charged particles in the magnetosphere.
The upper limit of \LX\ [0.5$-$10~keV] $\lesssim$ 5$\times$10$^{30}$~\UNITLUMI\ 
estimated from a fit of the spectrum above 2~keV assuming an additional power-law component with a fixed $\Gamma =$1.7
is similarly lower than is typical for middle-aged rotation-powered pulsars of a few $\times$10$^{5}$ year old \citep{Luca2005, Manzali2007}.
These results suggest that the neutron star originated in a supernova explosion
about 10$^6$ years ago.

The dip feature at $\sim$0.9 keV can be produced by fundamental cyclotron absorption by electrons
in the magnetic field with $\sim$7.5$\times$10$^{10}$~Gauss, which is typical of middle-aged radio pulsars \citep[e.g. ][]{Reisenegger2001},
or by protons in the magnetic fields with $\sim$10$^{14}$~Gauss, reminiscent of the X-ray Dim Isolated Neutron Star 
\citep[XDINS, see e.g.][]{Haberl2007,Kaplan2008}.
These stars have short pulsation periods of $\sim$a few hundred milli-second and/or small pulse fraction of $<$20\%.
The existing X-ray data sets do not have enough time resolution nor photon statistics to detect such a pulse.

Two other neutron stars have been discovered near other star forming regions \citep{Figer2005,Muno2006}.
Both of them have characteristics of magnetars --- neutron stars with strong magnetic fields 
up to 10$^{15}$~Gauss --- whose progenitor exploded within the last $\sim$10$^{4}$ years
and who might represent the high mass end of stars born in these star forming regions.
However, the neutron star discovered in the Carina Nebula cannot be co-eval with 
the current generation of stars.
Since the higher mass stars evolve faster,
the progenitor of the neutron star would have to be more massive than \etacar.
Eta Car's estimated initial mass $\gtrsim$150~\UNITSOLARMASS\ \citep{Hillier2001}
is much higher than the conventional progenitor mass of 
neutron stars (less than $\sim$25~\UNITSOLARMASS) and explosion of such a massive star would likely 
produce a black hole, instead of a neutron star 
although \citet{Woosley2002} found that explosions of stars with masses up to $\sim$60~\UNITSOLARMASS\ might
produce a neutron star under certain circumstances.

Based on standard stellar evolution models,
the progenitor of the neutron star should have been born 6$-$30$\times$10$^{6}$ year ago 
with an initial mass between 8$-$25~\UNITSOLARMASS.
This is significantly earlier than formation of the massive stars in the Trumpler 14 and 16 clusters
which occurred less than 3$\times$10$^{6}$ years ago
\citep{Degioia2001,Massey2001}.
This suggests that the Carina Nebula has experienced at least 2 episodes of star formation.
This is consistent with stellar population studies \citep{Degioia2001} which also suggest that intermediate-mass stars have
formed continuously over the last 10$^{7}$ years.
An alternative scenario is that EHG7 escaped from a nearby star forming region,
coming close to the Carina nebula by chance.
One possible site is the Carina flare region located at $\sim$2$-$10\DEGREE\ above the Carina nebula in Galactic coordinates.
This region is suspected to have experienced major star forming activity over 2$\times$10$^{7}$~yr \citep{Fukui1999}.
Another possibility is that \ess\ thermalizes through Bondi-Hoyle-Lyttleton accretion from 
the nebular gas \citep{Blaes1993}.
In that case, it can be much older than 10$^{6}$~yr and its birth place is unpredictable.

The last episode of star formation in the Carina Nebula appears to have been nearly
contemporaneous with the supernova explosion of the progenitor of this neutron star.
This may suggest that the expanding \ion{H}{2} region, wind-blown bubble and/or supernova explosion of this star 
(and perhaps others yet undetected) played a role in triggering the last episode of star formation.
Although the neutron star is currently well away from the nebula center,
it could have moved the projected distance in $\sim$10$^{6}$ year with a transverse velocity of only $\sim$6~\UNITVEL,
which is much lower than typical kick velocities of radio pulsars \cite[several hundred \UNITVEL, ][]{Lyne1994}.

Discovery of a neutron star supports the argument that the diffuse high energy emission observed in the Carina
Nebula originated in an ancient supernova explosion.
Some measurements require multiple supernova explosions in this field \citep{Yonekura2005,Hamaguchi2007a}, so that
the neutron star may be one of many neutron stars and black holes hidden in the Carina Nebula.
If the soft extended X-ray plasma in reality heated up by supernova explosions $\sim$10$^{6}$ years ago,
it would represent an important phase of the evolution of inter stellar medium when hot gas produced by 
supernova remnants merge together to form the hot ionized intercloud medium (HIM) component
\citep{McKee1977}.
This result may have implications for the origin of diffuse X-ray emission observed now from many star forming regions
\citep{Townsley2003,Gudel2008}.


\acknowledgments

We are grateful to T.R. Gull, K.E. Nielsen, S. Drake, K. Mukai, M. Ishida, C. Markwardt and anonymous referee for useful comments.
This work is performed while K.H. was supported by the NASA Astrobiology Program under CAN 03-OSS-02.
This research has made use of data and softwares obtained from the High Energy Astrophysics Science Archive Research Center (HEASARC), 
provided by NASA's Goddard Space Flight Center and Chandra X-ray Center (CXC).

{\it Facilities:} \facility{CXO (ACIS-I)}, \facility{XMM-Newton (EPIC)}, \facility{ROSAT (PSPC, HRI)}, \facility{EINSTEIN (HRI)}
\facility{Spitzer (IRAC)}, \facility{IRSF}

\bibliographystyle{apj}

\begin{thebibliography}{33}
\expandafter\ifx\csname natexlab\endcsname\relax\def\natexlab#1{#1}\fi

\bibitem[{{Albacete Colombo} {et~al.}(2003){Albacete Colombo}, {M{\'e}ndez}, \&
  {Morrell}}]{Colombo2003}
{Albacete Colombo}, J.~F., {M{\'e}ndez}, M., \& {Morrell}, N.~I. 2003, \mnras,
  346, 704

\bibitem[Blaes \& Madau(1993)]{Blaes1993} Blaes, O., \& Madau, P.\ 1993, \apj, 403, 690 

\bibitem[Beuermann et al.(2006)]{Beuermann2006} Beuermann, K., Burwitz, V., \& Rauch, T.\ 2006, \aap, 458, 541 

\bibitem[{{Broos} {et~al.}(2007){Broos}, {Feigelson}, {Townsley}, {Getman},
  {Wang}, {Garmire}, {Jiang}, \& {Tsuboi}}]{Broos2007}
{Broos}, P.~S., {Feigelson}, E.~D., {Townsley}, L.~K., {Getman}, K.~V., {Wang},
  J., {Garmire}, G.~P., {Jiang}, Z., \& {Tsuboi}, Y. 2007, \apjs, 169, 353

\bibitem[{{Davidson} \& {Humphreys}(1997)}]{Davidson1997}
{Davidson}, K., \& {Humphreys}, R.~M. 1997, \araa, 35, 1

\bibitem[{{De Luca} {et~al.}(2005){De Luca}, {Caraveo}, {Mereghetti},
  {Negroni}, \& {Bignami}}]{Luca2005}
{De Luca}, A., {Caraveo}, P.~A., {Mereghetti}, S., {Negroni}, M., \& {Bignami},
  G.~F. 2005, \apj, 623, 1051

\bibitem[{{DeGioia-Eastwood} {et~al.}(2001){DeGioia-Eastwood}, {Throop},
  {Walker}, \& {Cudworth}}]{Degioia2001}
{DeGioia-Eastwood}, K., {Throop}, H., {Walker}, G., \& {Cudworth}, K.~M. 2001,
  \apj, 549, 578

\bibitem[{{Elmegreen}(1998)}]{Elmegreen1998}
{Elmegreen}, B.~G. 1998, in Astronomical Society of the Pacific Conference
  Series, Vol. 148, Origins, ed. C.~E. {Woodward}, J.~M. {Shull}, \& H.~A.
  {Thronson}, Jr., 150

\bibitem[{{Ezoe} {et~al.}(2008){Ezoe}, {Hamaguchi}, {Gruendl}, {Chu}, {Petre},
  \& {Corcoran}}]{Ezoe2008}
{Ezoe}, Y., {Hamaguchi}, K., {Gruendl}, R.~A., {Chu}, Y.-H., {Petre}, R., \&
  {Corcoran}, M.~F. 2008, PASJ, in print (astro-ph/0809.3495)

\bibitem[{{Fazio} {et~al.}(2004){Fazio et al.}}]{Fazio2004}
{Fazio}, G.~G., and 64 others, 2004, \apjs, 154, 10

\bibitem[{{Figer} {et~al.}(2005){Figer}, {Najarro}, {Geballe}, {Blum}, \&
  {Kudritzki}}]{Figer2005}
{Figer}, D.~F., {Najarro}, F., {Geballe}, T.~R., {Blum}, R.~D., \& {Kudritzki},
  R.~P. 2005, \apjl, 622, L49

\bibitem[{{Fukui} {et~al.}(1999){Fukui}, {Onishi}, {Abe}, {Kawamura},
  {Tachihara}, {Yamaguchi}, {Mizuno}, \& {Ogawa}}]{Fukui1999}
{Fukui}, Y., {Onishi}, T., {Abe}, R., {Kawamura}, A., {Tachihara}, K.,
  {Yamaguchi}, R., {Mizuno}, A., \& {Ogawa}, H. 1999, \pasj, 51, 751

\bibitem[{{Gagn\'{e}} {et~al.}(1995){Gagn\'{e}}, {Caillault}, \&
  {Stauffer}}]{Gagne1995}
{Gagn\'{e}}, M., {Caillault}, J.~P., \& {Stauffer}, J.~R. 1995, \apj, 445, 280

\bibitem[{{G{\"u}del} {et~al.}(2008){G{\"u}del}, {Briggs}, {Montmerle},
  {Audard}, {Rebull}, \& {Skinner}}]{Gudel2008}
{G{\"u}del}, M., {Briggs}, K.~R., {Montmerle}, T., {Audard}, M., {Rebull}, L.,
  \& {Skinner}, S.~L. 2008, Science, 319, 309

\bibitem[Haberl(2007)]{Haberl2007} Haberl, F.\ 2007, \apss, 308, 181

\bibitem[{{Hamaguchi} {et~al.}(2007){Hamaguchi}, {Petre}, {Matsumoto},
  {Tsujimoto}, {Holt}, {Ezoe}, {Ozawa}, {Tsuboi}, {Soong}, {Kitamoto},
  {Sekiguchi}, \& {Kokubun}}]{Hamaguchi2007a}
{Hamaguchi}, K., and 11 others, 2007, \pasj, 59, 151

\bibitem[{{Harnden} {et~al.}(1984){Harnden}, {Fabricant}, {Harris}, \&
  {Schwarz}}]{Harnden1984}
{Harnden}, Jr., F.~R., {Fabricant}, D.~G., {Harris}, D.~E., \& {Schwarz}, J.
  1984, SAO Special Report, 393

\bibitem[{{Hillier} {et~al.}(2001){Hillier}, {Davidson}, {Ishibashi}, \&
  {Gull}}]{Hillier2001}
{Hillier}, D.~J., {Davidson}, K., {Ishibashi}, K., \& {Gull}, T. 2001, \apj,
  553, 837

\bibitem[Kahabka \& van den Heuvel(1997)]{Kahabka1997} Kahabka, P., \& van den Heuvel, E.~P.~J.\ 1997, \araa, 35, 69 

\bibitem[{{Kalberla} {et~al.}(2005){Kalberla}, {Burton}, {Hartmann}, {Arnal},
  {Bajaja}, {Morras}, \& {P{\"o}ppel}}]{Kalberla2005}
{Kalberla}, P.~M.~W., {Burton}, W.~B., {Hartmann}, D., {Arnal}, E.~M.,
  {Bajaja}, E., {Morras}, R., \& {P{\"o}ppel}, W.~G.~L. 2005, \aap, 440, 775

\bibitem[Kaplan(2008)]{Kaplan2008} Kaplan, D.~L.\ 2008, 40 Years of Pulsars: Millisecond Pulsars, Magnetars and More, 
ed. C.~{Bassa}, Z.~{Wang}, A.~{Cumming}, \& V.~M.  {Kaspi}, 983, 331

\bibitem[{{Leutenegger} {et~al.}(2003){Leutenegger}, {Kahn}, \&
  {Ramsay}}]{Leutenegger2003}
{Leutenegger}, M.~A., {Kahn}, S.~M., \& {Ramsay}, G. 2003, \apj, 585, 1015

\bibitem[{{Lyne} \& {Lorimer}(1994)}]{Lyne1994}
{Lyne}, A.~G., \& {Lorimer}, D.~R. 1994, \nat, 369, 127

\bibitem[{{Manchester} {et~al.}(2005){Manchester}, {Hobbs}, {Teoh}, \&
  {Hobbs}}]{Manchester2005}
{Manchester}, R.~N., {Hobbs}, G.~B., {Teoh}, A., \& {Hobbs}, M. 2005, \aj, 129,
  1993

\bibitem[{{Manzali} {et~al.}(2007){Manzali}, {De Luca}, \&
  {Caraveo}}]{Manzali2007}
{Manzali}, A., {De Luca}, A., \& {Caraveo}, P.~A. 2007, \apj, 669, 570

\bibitem[{{Massey} {et~al.}(2001){Massey}, {DeGioia-Eastwood}, \&
  {Waterhouse}}]{Massey2001}
{Massey}, P., {DeGioia-Eastwood}, K., \& {Waterhouse}, E. 2001, \aj, 121, 1050

\bibitem[{{McClure-Griffiths et al.}(2005){McClure-Griffiths et al.}}]{McClure2005} 
McClure-Griffiths, N.~M., Dickey, J.~M., Gaensler, B.~M., Green, A.~J., 
Haverkorn, M., \& Strasser, S.\ 2005, \apjs, 158, 178

\bibitem[{{McKee} \& {Ostriker}(1977)}]{McKee1977}
{McKee}, C.~F., \& {Ostriker}, J.~P. 1977, \apj, 218, 148

\bibitem[{{Muno} {et~al.}(2006){Muno}, {Clark}, {Crowther}, {Dougherty}, {de
  Grijs}, {Law}, {McMillan}, {Morris}, {Negueruela}, {Pooley}, {Portegies
  Zwart}, \& {Yusef-Zadeh}}]{Muno2006}
{Muno}, M.~P., and 11 others, 2006, \apjl, 636, L41
  
\bibitem[{{Pires} \& {Motch}(2008)}]{Pires2008}
{Pires}, A.~M., \& {Motch}, C. 2008, in American Institute of Physics
  Conference Series, Vol. 983, 40 Years of Pulsars: Millisecond Pulsars,
  Magnetars and More, ed. C.~{Bassa}, Z.~{Wang}, A.~{Cumming}, \& V.~M.
  {Kaspi}, 363--365

\bibitem[{{Pires} {et~al.}(2008)}]{Pires2008b}
{Pires}, A.~M., {Motch}, C., {Turolla}, R., {Treves}, A., \& {Popov}, S.~B. 2008, \aap, accepted (astro-ph/0812.4151)

\bibitem[{{Reisenegger}(2001)}]{Reisenegger2001}
{Reisenegger}, A. 2001, in Astronomical Society of the Pacific Conference
  Series, Vol. 248, Magnetic Fields Across the Hertzsprung-Russell Diagram, ed.
  G.~{Mathys}, S.~K. {Solanki}, \& D.~T. {Wickramasinghe}, 469

\bibitem[Sala \& Hernanz(2005)]{Sala2005} Sala, G., \& Hernanz, M.\ 2005, \aap, 439, 1061 

\bibitem[{{Sanchawala} {et~al.}(2007{\natexlab{a}}){Sanchawala}, {Chen}, {Lee},
  {Chu}, {Nakajima}, {Tamura}, {Baba}, \& {Sato}}]{Sanchawala2007a}
{Sanchawala}, K., {Chen}, W.-P., {Lee}, H.-T., {Chu}, Y.-H., {Nakajima}, Y.,
  {Tamura}, M., {Baba}, D., \& {Sato}, S. 2007{\natexlab{a}}, \apj, 656, 462

\bibitem[{{Sanchawala} {et~al.}(2007{\natexlab{b}}){Sanchawala}, {Chen},
  {Ojha}, {Ghosh}, {Nakajima}, {Tamura}, {Baba}, {Sato}, \&
  {Tsujimoto}}]{Sanchawala2007b}
{Sanchawala}, K., and 8 others, 2007{\natexlab{b}}, \apj, 667, 963
  
\bibitem[{{Sanwal} {et~al.}(2002){Sanwal}, {Pavlov}, {Zavlin}, \&
  {Teter}}]{Sanwal2002}
{Sanwal}, D., {Pavlov}, G.~G., {Zavlin}, V.~E., \& {Teter}, M.~A. 2002, \apjl,
  574, L61

\bibitem[{{Seward} {et~al.}(1979){Seward}, {Forman}, {Giacconi}, {Griffiths},
  {Harnden}, {Jones}, \& {Pye}}]{Seward1979}
{Seward}, F.~D., {Forman}, W.~R., {Giacconi}, R., {Griffiths}, R.~E.,
  {Harnden}, F. R.~J., {Jones}, C., \& {Pye}, J.~P. 1979, \apjl, 234, L55

\bibitem[{{Smith}(2006)}]{Smith2006}
{Smith}, N. 2006, \mnras, 367, 763

\bibitem[{{Smith} {et~al.}(2000){Smith}, {Egan}, {Carey}, {Price}, {Morse}, \&
  {Price}}]{Smith2000}
{Smith}, N., {Egan}, M.~P., {Carey}, S., {Price}, S.~D., {Morse}, J.~A., \&
  {Price}, P.~A. 2000, \apjl, 532, L145

\bibitem[{{Str{\" u}der} {et~al.}(2001){Str{\" u}der}, {Briel}, {Dennerl},
  {Hartmann}, {Kendziorra}, {Meidinger}, {Pfeffermann}, {Reppin}, {Aschenbach},
  {Bornemann}, {Br{\" a}uninger}, {Burkert}, {Elender}, {Freyberg}, {Haberl},
  {Hartner}, {Heuschmann}, {Hippmann}, {Kastelic}, {Kemmer}, {Kettenring},
  {Kink}, {Krause}, {M{\" u}ller}, {Oppitz}, {Pietsch}, {Popp}, {Predehl},
  {Read}, {Stephan}, {St{\" o}tter}, {Tr{\" u}mper}, {Holl}, {Kemmer},
  {Soltau}, {St{\" o}tter}, {Weber}, {Weichert}, {von Zanthier},
  {Carathanassis}, {Lutz}, {Richter}, {Solc}, {B{\" o}ttcher}, {Kuster},
  {Staubert}, {Abbey}, {Holland}, {Turner}, {Balasini}, {Bignami}, {La
  Palombara}, {Villa}, {Buttler}, {Gianini}, {Lain{\' e}}, {Lumb}, \&
  {Dhez}}]{Struder2001}
{Str{\" u}der}, L., et al. 2001, \aap, 365, L18

\bibitem[{{Townsley}(2006)}]{Townsley2006}
{Townsley}, L.~K. 2006, in the STScI May Symposium, "Massive Stars: From Pop
  III and GRBs to the Milky Way, ed. M.~Livio, (astro--ph/0608173)

\bibitem[{{Townsley} {et~al.}(2003){Townsley}, {Feigelson}, {Montmerle},
  {Broos}, {Chu}, \& {Garmire}}]{Townsley2003}
{Townsley}, L.~K., {Feigelson}, E.~D., {Montmerle}, T., {Broos}, P.~S., {Chu},
  Y.-H., \& {Garmire}, G.~P. 2003, \apj, 593, 874

\bibitem[{{Tsuruta}(2006)}]{Tsuruta2006}
{Tsuruta}, S. 2006, in American Institute of Physics Conference Series, Vol.
  847, Origin of Matter and Evolution of Galaxies, ed. S.~{Kubono}, W.~{Aoki},
  T.~{Kajino}, T.~{Motobayashi}, \& K.~{Nomoto}, 163--170

\bibitem[Woods \& Thompson(2006)]{Woods2006} Woods, P.~M., \& Thompson, C.\ 2006, 
in Compact stellar X-ray sources, ed. W. H. G. Lewin \& M. van der Klis (Cambridge: Cambridge Univ. Press)

\bibitem[{{Woosley} {et~al.}(2002){Woosley}, {Heger}, \&
  {Weaver}}]{Woosley2002}
{Woosley}, S.~E., {Heger}, A., \& {Weaver}, T.~A. 2002, Reviews of Modern
  Physics, 74, 1015

\bibitem[{{Yonekura} {et~al.}(2005){Yonekura}, {Asayama}, {Kimura}, {Ogawa},
  {Kanai}, {Yamaguchi}, {Barnes}, \& {Fukui}}]{Yonekura2005}
{Yonekura}, Y., {Asayama}, S., {Kimura}, K., {Ogawa}, H., {Kanai}, Y.,
  {Yamaguchi}, N., {Barnes}, P.~J., \& {Fukui}, Y. 2005, \apj, 634, 476

\end{thebibliography}

\end{document}